\documentclass[manuscript]{acmart}

\AtBeginDocument{%
  }

\setcopyright{acmlicensed}
\copyrightyear{2024}
\acmYear{2024}
\acmDOI{XXXXXXX.XXXXXXX}

\acmConference[ReAI '24]{1st Workshop on Responsible Artificial Intelligence}{September 11--13, 2024}{Piraeus, Greece}
\acmISBN{978-1-4503-XXXX-X/18/06}

\begin{document}

\title{Artificial Intelligence in Education: Ethical Considerations and Insights from Ancient Greek Philosophy}

\author{Kostas Karpouzis}
\orcid{0000-0002-4615-6751}
\email{kkarpou@panteion.gr}
\affiliation{%
  \institution{Department of Communication, Media and Culture, Panteion University of Social and Political Sciences
  \city{Athens}
  \country{Greece}}
}

\renewcommand{\shortauthors}{K. Karpouzis}

\begin{abstract}
This paper explores the ethical implications of integrating Artificial Intelligence (AI) in educational settings, from primary schools to universities, while drawing insights from ancient Greek philosophy to address emerging concerns. As AI technologies increasingly influence learning environments, they offer novel opportunities for personalized learning, efficient assessment, and data-driven decision-making. However, these advancements also raise critical ethical questions regarding data privacy, algorithmic bias, student autonomy, and the changing roles of educators. This research examines specific use cases of AI in education, analyzing both their potential benefits and drawbacks. By revisiting the philosophical principles of ancient Greek thinkers such as Socrates, Aristotle, and Plato, we discuss how their writings can guide the ethical implementation of AI in modern education. The paper argues that while AI presents significant challenges, a balanced approach informed by classical philosophical thought can lead to an ethically sound transformation of education. It emphasizes the evolving role of teachers as facilitators and the importance of fostering student initiative in AI-rich environments.

\end{abstract}


\maketitle

\section{Introduction}
The integration of Artificial Intelligence (AI) in education has been rapidly accelerating, transforming traditional learning environments and methodologies \cite{holmes2019artificial}. From personalized learning platforms \cite{mavrikis2019towards} to automated grading systems, AI is reshaping how knowledge is transmitted, acquired, and assessed in both formal and informal educational settings \cite{zawacki2019systematic}.

While the potential benefits of AI in education are significant, including enhanced personalization and improved accessibility \cite{luckin2016intelligence}, the increasing reliance on these technologies raises important ethical concerns \cite{selwyn2019should}. Issues such as data privacy, algorithmic bias, and the impact on human agency in learning have become pressing topics of discussion among educators, policymakers, and ethicists \cite{regan2019ethical}.
Interestingly, as we grapple with these thoroughly modern challenges, ancient Greek philosophy offers a valuable lens through which to examine and address these ethical dilemmas (\cite{vallor2016technology}, \cite{karpouzis2024plato}). The teachings of Socrates, Plato, and Aristotle, among others, provide enduring insights into the nature of knowledge, ethics, and human development that remain relevant in our AI-driven world \cite{kristjansson2015aristotelian}.

This paper aims to explore the ethical implications of AI in education, drawing upon the wisdom of ancient Greek philosophy to guide our approach to these contemporary issues. By bridging classical thought with modern technology, we seek to develop a framework for the ethical implementation of AI in educational contexts that balances innovation with human values \cite{pena2019robot}. Our discussion will encompass the current state of AI in education, key ethical issues, relevant insights from Greek philosophy, the evolving role of teachers, and the importance of fostering student autonomy. Through this exploration, we aim to present a critical yet optimistic view of AI's potential to revolutionize education when guided by sound ethical principles.

\section{The Current State of AI in Education: Use Cases and Implications}
The integration of Artificial Intelligence (AI) in educational settings has been rapidly evolving, offering innovative solutions to longstanding challenges in teaching and learning \cite{holmes2019artificial}. As AI technologies continue to affect various aspects of education, from primary schools to higher education institutions, they are reshaping traditional paradigms and opening new avenues for enhanced learning experiences \cite{zawacki2019systematic}.

Integration of AI concepts in education is typically aimed at enhancing learning experiences, streamlining administrative processes, and providing data-driven insights for educational decision-making \cite{luckin2016intelligence}. These applications span a wide spectrum, each with its own set of potential benefits and challenges. One of the most prominent use cases of AI in education is personalized learning platforms. These systems leverage AI algorithms to adapt content and pacing to individual student needs, creating tailored learning experiences that cater to diverse learning styles and abilities (\cite{mavrikis2019towards}, \cite{holmes2019artificial}). By analyzing readily available, vast amounts of data on student performance and behavior, these platforms can identify knowledge gaps, suggest appropriate textual, audiovisual or interactive resources, and adjust the difficulty level of content in real-time. This level of personalization has the potential to significantly improve learning outcomes by ensuring that each student receives instruction that is optimally challenging and engaging.

Another significant application of AI in education is automated grading and feedback systems. These tools are designed to assess student work and provide timely feedback, potentially reducing the workload on educators and allowing them to focus on more complex aspects of teaching \cite{balfour2013automated}. While particularly useful for objective assessments, recent advancements have also shown promise in evaluating more subjective work, such as essays. However, the use of such systems also raises questions about the nature of assessment and the role of human judgment in evaluating student work.

AI-powered tutoring and virtual assistants represent another frontier in educational technology. These intelligent tutoring systems provide on-demand support to students, offering explanations, answering questions, and guiding learners through complex problem-solving processes \cite{kulik2016effectiveness}, supplementing traditional classroom instruction, and providing additional support outside of school hours. However, the effectiveness of these systems compared to human tutors, and their impact on student-teacher relationships, remain topics of ongoing research and debate, especially when it comes to establishing rapport with students.

In the field of educational administration, AI is being employed for predictive analytics and resource allocation. Predictive models can identify at-risk students and suggest interventions, potentially improving retention rates and academic outcomes \cite{tsai2019complexity}. Similarly, AI systems are being used to optimize scheduling, resource distribution, and other administrative tasks, potentially leading to more efficient and effective school operations \cite{popenici2017exploring}.

While these applications of AI in education offer significant potential benefits, they also raise important concerns and challenges. Privacy issues related to the collection and analysis of student data are paramount \cite{regan2019ethical}. The vast amount of data required for AI systems to function effectively poses risks to student privacy and raises questions about data ownership, storage, and use. Moreover, there is growing awareness of the potential for bias in algorithmic decision-making. AI systems, if not carefully designed and monitored, could perpetuate or exacerbate existing inequalities in education (\cite{luan2020challenges}, \cite{karpouzis2024plato}). This concern extends to issues of equity in access to AI-enhanced education, as disparities in technological resources could further widen the educational gap between privileged and underprivileged students \cite{reich2017good}.

The increasing reliance on AI in education also raises concerns about the potential diminishment of human interaction in the learning process. While AI can, indeed, provide valuable support and personalization, the role of human teachers in fostering critical thinking, creativity, and social-emotional skills remains crucial (\cite{selwyn2019should}, \cite{cowie2011emotion}). Striking the right balance between AI-driven instruction and human-led teaching represents a significant challenge for educators and policymakers alike. Similarly, the impact of AI on academic integrity also constitutes an emerging concern, since the development of sophisticated AI writing tools makes ensuring the authenticity of student work \cite{stoesz2019academic} a challenging effort. In this context, both educators and educational institutions are grappling with how to adapt their policies and practices to address these new forms of potential academic misconduct.

Regulatory frameworks meant to address these unique challenges are currently discussed or are already in place. The EU AI Act, for instance, classifies certain AI systems used in educational or vocational training as high-risk, requiring them to meet specific requirements before deployment \cite{veale2021demystifying}. This legislation reflects growing recognition of the potential impacts of AI in education and the need for robust governance frameworks. Data protection regulations, such as the General Data Protection Regulation (GDPR) in the European union, have significant implications for the collection and processing of student data in AI systems (\cite{slade2013learning}, \cite{panagopoulou2023legal}. These regulations require educational institutions and technology providers to implement strong data protection measures and ensure transparency in data usage. On a global scale, initiatives such as UNESCO's recommendations on AI ethics provide a framework for the ethical use of AI in education \cite{morandin2023ten}. These guidelines emphasize the importance of human rights, inclusion, and transparency in the development and deployment of AI technologies in educational contexts.

However, regulating AI in education presents unique challenges. There is a delicate balance to be struck between fostering innovation and protecting student interests \cite{tuomi2019impact}. Overly restrictive regulations could stifle the development of beneficial AI technologies, while insufficient oversight could lead to negative consequences for students and educators. In addition, the global nature of online learning and educational technology further complicates regulatory efforts. AI-powered online learning platforms often operate across national borders, raising questions about jurisdiction and the applicability of national regulations \cite{zuboff2023age} and developing international standards and cooperation mechanisms will be crucial in addressing these cross-border challenges. Another important issue has to do with the rapid pace of technological advancement in AI: regulations must be flexible enough to accommodate new developments while still providing meaningful protection and guidance \cite{cath2018artificial} and this requires ongoing dialogue between policymakers, educators, technologists, and other stakeholders to ensure that regulatory approaches remain relevant and effective.

Given these challenges, it is clear that the integration of AI in education holds great promise, but may also bring in significant risks. Realizing the potential of AI to enhance learning outcomes, increase educational access, and improve administrative efficiency will require careful consideration of ethical implications, thoughtful policy development, and ongoing evaluation of its impacts. The future of AI in education will likely be shaped by our ability to harness its benefits while effectively addressing its risks and challenges.

\section{Key Ethical Issues}
The integration of Artificial Intelligence (AI) in education, while offering numerous benefits, also presents a number of ethical challenges that demand careful consideration. These ethical issues touch upon fundamental aspects of education, privacy, equity, and human development, requiring a nuanced approach to ensure that the implementation of AI in educational settings aligns with our core values and educational goals.

One of the primary ethical concerns in AI-driven education is the issue of data privacy and security. The effectiveness of AI systems in personalizing learning experiences and providing insights into student performance relies heavily on the collection and analysis of vast amounts of student data (\cite{regan2019ethical}, \cite{tsatiris2021developing}). This data can include everything from academic performance and learning patterns to behavioral information and even biometric or medical data in some cases. The extensive data collection raises significant privacy concerns, particularly given that much of this data pertains to minors who may not fully understand the implications of their data being collected and analyzed \cite{prinsloo2017ethics}.

The potential for data breaches or misuse of student information is a serious concern. Educational institutions and technology providers must implement robust security measures to protect sensitive student data. Moreover, there are ethical questions (beyond regulatory demands, such as GDPR) about the long-term storage and potential future or dual uses of this data: could information collected about a student's learning difficulties or behavioral patterns be used in ways that may disadvantage them later in life? The principle of data minimization – collecting only the data necessary for specific educational purposes – becomes crucial in this context \cite{slade2013learning}.

Another significant ethical issue is that of algorithmic bias and fairness. AI systems are only as unbiased as the data they are trained on and the humans who design them. In the context of education, where AI is increasingly being used for high-stakes decisions such as college admissions or course recommendations, the potential for bias can have far-reaching consequences \cite{baker2009}. Biases in AI systems could perpetuate or even exacerbate existing inequalities in education, disadvantaging students from minority groups or low-income backgrounds \cite{panagopoulou2023legal}. For instance, if an AI system used for college admissions is trained on historical data that reflects past discriminatory practices, it may perpetuate these biases in its recommendations. Similarly, AI-powered adaptive learning systems might make assumptions about a student's abilities or potential based on demographic information, potentially limiting their educational opportunities \cite{reich2017good}. Ensuring fairness and mitigating bias in AI systems is not just a technical challenge but an ethical imperative in educational contexts.

The use of AI in education also raises important questions about autonomy and human agency in learning. While personalized learning powered by AI can adapt to individual student needs, there is a risk of over-reliance on algorithmic recommendations. This could potentially limit students' ability to explore diverse subjects, take intellectual risks, or develop critical thinking skills \cite{selwyn2019should}. The ethical challenge lies in balancing the benefits of AI-driven personalization with the need to foster independent thinking and decision-making skills in students. Moreover, the increasing use of AI in education may impact the development of important social and emotional skills. Human interaction plays a crucial role in education, not just in transmitting knowledge but in modeling behaviors, fostering empathy, and developing social skills. As AI takes on more roles in the educational process, from tutoring to assessment, there is a risk of diminishing these vital human elements of education \cite{pena2019robot}. Ethically, we must consider how to leverage AI in ways that enhance rather than replace meaningful human interactions in educational settings.

The issues of transparency and explainability in AI systems used in education are also relevant here. Many AI algorithms, particularly those using deep learning techniques, operate as ``black boxes'', making decisions or recommendations in ways that are not easily interpretable by humans \cite{rudin2019stop}. In an educational context, where decisions can significantly impact a student's future, the lack of transparency in AI decision-making processes is problematic. Students, parents, and educators have a right to understand how educational decisions are being made, especially when AI systems play a role in these decisions.

The authenticity of AI-assisted work presents another ethical challenge. As AI tools become more sophisticated in generating text, solving problems, and even creating art, questions arise about the boundaries of acceptable use in educational settings \cite{stoesz2019academic}. How do we define originality and authorship in an era where AI can produce human-like text or solve complex problems? This issue touches on fundamental questions of academic integrity and the very purpose of education itself.

Finally, equity in access to AI-enhanced education is a critical ethical consideration. While AI has the potential to democratize education by providing personalized learning experiences at scale, there is also a risk of creating or widening digital divides \cite{holmes2019artificial}. Students from disadvantaged backgrounds may have limited access to AI-powered educational tools, potentially falling further behind their more privileged peers. Ensuring equitable access to AI-enhanced education is not just a logistical challenge but an ethical imperative in promoting educational equality.

Addressing these ethical challenges requires a multifaceted approach involving educators, policymakers, technologists, and ethicists. It calls for the development of ethical guidelines and governance frameworks specifically tailored to the use of AI in education. These frameworks must be flexible enough to adapt to rapidly evolving technology while providing robust protection for students' rights and well-being. Moreover, fostering AI literacy among educators, students, and parents is crucial. Understanding the capabilities, limitations, and ethical implications of AI is becoming an essential skill in the modern world. By promoting this understanding, we can empower stakeholders to make informed decisions about the use of AI in education and to engage critically with these technologies \cite{zuboff2023age}.

\section{Ancient Greek Philosophy and AI Ethics}
The rapid advancement of Artificial Intelligence (AI) in education presents ethical challenges that, while seemingly novel, touch upon fundamental questions of knowledge, virtue, and human nature that have been explored for millennia. Ancient Greek philosophy, with its rich tradition of ethical inquiry and educative theory, offers a valuable lens through which to examine and address these modern dilemmas. By revisiting the ideas of Socrates, Plato, Aristotle, and the Stoics, we can gain insights into the ethical implementation of AI in educational contexts \cite{karpouzis2024would}.

The \emph{Socratic} method, developed by Socrates and documented by Plato, provides a powerful framework for considering the role of AI in fostering critical thinking. Central to this method is the practice of systematic questioning to stimulate critical thinking and illuminate ideas \cite{vlastos1991socrates}. In the context of AI-driven education, we must consider how to preserve and enhance this aspect of learning. While AI can efficiently deliver information and even adapt to individual learning styles, it may struggle to replicate the nuanced, dynamic process of Socratic dialogue; the challenge, then, is to design AI systems that not only provide information but also encourage questioning and critical analysis. Some researchers have attempted to develop AI tutors that emulate Socratic questioning \cite{olney2012guru}, but these systems often lack the sophistication and adaptability of human interlocutors. The Socratic ideal reminds us that education is not merely about transferring knowledge, but about developing the capacity for independent, critical thought. As we integrate AI into education, we must ensure that it enhances rather than diminishes this crucial aspect of learning.

Plato's \emph{Theory of Forms}, which posits that abstract ideas have a real, perfect existence beyond the physical world, offers another interesting perspective on AI in education. In Plato's view, true knowledge involves grasping these perfect forms, rather than merely understanding their imperfect physical manifestations \cite{kraut2017plato}. This concept raises intriguing questions about the nature of knowledge in an AI-driven educational landscape: can AI, which operates on data and algorithms, help students access these deeper, abstract truths or does the data-driven nature of AI limit its ability to engage with the kind of abstract thinking Plato valued? Moreover, Plato's \emph{Allegory of the Cave} \cite{karpouzis2024plato}, which illustrates the journey from ignorance to knowledge, takes on new relevance in the age of AI. Just as the prisoners in Plato's cave mistake shadows for reality, we must be cautious about mistaking AI-generated information or recommendations for ultimate truth. This allegory underscores the importance of developing AI literacy among students and educators, enabling them to critically evaluate AI-generated content and understand its limitations.

Aristotle's \emph{virtue} ethics emphasizes the development of good character traits or virtues, providing another valuable framework for considering AI in education. Aristotle believed that virtues are developed through practice and habit, not simply through acquiring knowledge \cite{kristjansson2015aristotelian}. This perspective raises important questions about the role of AI in character education. While AI can efficiently deliver information and even personalize learning experiences, can it contribute to the development of virtues like courage, temperance, or justice? Furthermore, Aristotle's concept of \emph{phronesis}, or practical wisdom, is particularly relevant to AI ethics in education. Phronesis involves the ability to make good judgments in complex, context-dependent situations \cite{dunne1993back}. As we increasingly rely on AI for decision-making in educational contexts, from personalized learning paths to college admissions, we must consider how to preserve and cultivate this kind of practical wisdom. AI systems, no matter how sophisticated, may struggle to replicate the nuanced, context-sensitive judgments that characterize phronesis.
The Stoic philosophy, with its emphasis on logic, ethics, and the cultivation of virtue, also offers valuable insights for AI ethics in education. The Stoic ideal of maintaining equanimity in the face of external circumstances resonates with the need to maintain human agency in an AI-driven world \cite{sellars2014stoicism}. As AI systems become more prevalent in education, the Stoic emphasis on internal locus of control reminds us of the importance of fostering students' sense of autonomy and self-efficacy. Moreover, the Stoic concept of \emph{oikeiosis}, or the process of coming to understand one's place in the broader community and cosmos, takes on new significance in the age of AI. As AI reshapes the educational landscape, we must consider how to help students understand their place in a world increasingly mediated by intelligent machines. This involves not only teaching technical skills but also fostering a broader understanding of the ethical and societal implications of AI.
Finally, the ancient Greek concept of \emph{paideia}, which encompasses the holistic education and cultural formation of the individual, provides an overarching framework for considering AI in education. Paideia involved not just the transmission of knowledge or formal education, but the cultivation of the whole person - intellectually, morally, and civically \cite{jaeger1939ideals}. As we integrate AI into education, we must ensure that we do not lose sight of this holistic vision of education. AI should be a tool for enhancing paideia, not a replacement for the rich, multifaceted process of human development that the ancient Greeks envisioned.

In considering these ancient philosophical concepts, we are reminded that education is not merely about the efficient transfer of information or the development of specific skills. It is about the formation of the whole person, the cultivation of wisdom and virtue, and the development of critical thinking skills. The insights from ancient Greek philosophy also underscore the importance of transparency and explainability in AI systems used in education. Just as Socrates insisted on clear definitions and logical reasoning, we should look for and demand clarity and understandability from our AI systems. This aligns with contemporary calls for explainable AI, particularly in high-stakes domains like education (\cite{rudin2019stop}, \cite{karpouzis2023explainable}). Furthermore, the ancient Greek emphasis on dialogue and dialectic reminds us of the importance of ongoing ethical deliberation as we integrate AI into education. We must foster continuous dialogue among educators, technologists, policymakers, and philosophers to navigate the complex ethical terrain of AI in education. This interdisciplinary approach, combining technological expertise with philosophical insight, echoes the ancient Greek ideal of a well-rounded education.

\section{The Evolving Role of Teachers}
The integration of Artificial Intelligence (AI) in education is reshaping the role of teachers in ways that echo ancient Greek philosophical concepts about education and the nature of knowledge. By examining this evolution through the lens of classical thought, we can gain valuable insights into how to navigate this transformation while preserving the essential human elements of teaching.

The shifting role of teachers from primary knowledge transmitters to learning facilitators mirrors the Socratic concept of the teacher as a ``midwife'' to knowledge. Socrates, as portrayed in Plato's dialogues, saw his role not as imparting information, but as helping students give birth to their own ideas through questioning and dialogue \cite{vlastos1991socrates}. In the age of AI, where information is readily accessible and AI systems can deliver personalized content, teachers are increasingly adopting this Socratic role. For instance, AI systems can analyze student performance data to identify areas of struggle, but it's the teacher who must engage in the Socratic process of questioning and guiding students to deeper understanding. A study by Holstein et al. \cite{holstein2018student} found that when teachers were provided with AI-generated insights about student performance, they were able to provide more targeted interventions. This exemplifies how AI can enhance the teacher's ability to practice the Socratic method more effectively.

The concept of \emph{paideia}, the holistic education and cultural formation central to ancient Greek thought, becomes even more relevant as AI takes over routine educational tasks. As mentioned, paideia involved not just the transmission of knowledge, but the cultivation of the whole person - intellectually, morally, and civically \cite{jaeger1939ideals}. As AI systems handle content delivery and basic assessment, teachers are freed to focus on this broader, more holistic aspect of education. This shift aligns with Aristotle's concept of phronesis or practical wisdom. Aristotle believed that true education involved not just knowledge (\emph{episteme}) or technical skill (\emph{techne}), but the ability to make good judgments in complex, context-dependent situations \cite{dunne1993back}. In an AI-driven educational landscape, the teacher's role in cultivating this phronesis becomes even more crucial. While AI can provide information and even adapt to learning styles, it cannot replicate the nuanced, context-sensitive judgments that characterize phronesis.

Moreover, the teacher's evolving role as a curator of learning experiences and a guide in ethical reasoning reflects Plato's allegory of the cave. Just as the philosopher in Plato's allegory guides others towards true understanding beyond mere shadows \cite{kraut2017plato}, teachers in the AI era must help students navigate beyond the surface-level information provided by AI systems to deeper, more critical understanding.

The need for teachers to develop AI literacy and critical evaluation skills echoes the ancient Greek emphasis on logos or rational discourse. The Sophists, despite their controversial reputation, recognized the importance of teaching critical thinking and the ability to argue different perspectives \cite{guthrie1971sophists}. In the context of AI-driven education, teachers must not only understand AI systems but also teach students to critically evaluate AI-generated content and recommendations.

The shift towards more collaborative and interdisciplinary approaches in teaching reflects the ancient Greek ideal of the academy. Plato's Academy, for instance, was a place of collaborative inquiry where various disciplines were studied in relation to each other \cite{dillon2003heirs}. In the AI era, teachers are increasingly working in teams that may include data scientists and AI specialists, echoing this interdisciplinary approach to knowledge creation and dissemination. 

However, this evolution is not without challenges. The risk of over-reliance on AI systems resonates with Plato's concerns about the written word in the Phaedrus dialogue. Just as Plato worried that reliance on writing would weaken memory and true understanding \cite{kraut2017plato}, we must be cautious about over-reliance on AI at the expense of human judgment and intuition in education. Furthermore, the need to redefine teacher education in light of AI echoes the ancient Greek concern with the proper training of educators. Plato's Republic, for instance, devotes considerable attention to the education of the guardians who would be responsible for educating others \cite{kraut2017plato}. In our context, this translates to the need for teacher training programs that prepare educators for the AI-enhanced landscape, incorporating instruction on AI literacy and effective integration of AI tools in pedagogical practice.

\section{Fostering Student Initiative and Autonomy}
The integration of AI in education presents both opportunities and challenges for fostering student initiative and autonomy. By examining this issue through the lens of ancient Greek philosophy, we can gain valuable insights into how to promote self-directed learning and independent thinking in an AI-rich educational environment.

The concept of student autonomy resonates strongly with the Socratic method, which aimed to cultivate independent thought through rigorous questioning. Socrates famously claimed that he was not a teacher, but rather a midwife helping others give birth to their own ideas \cite{vlastos1991socrates}. In the context of AI-enhanced education, we must consider how to preserve and enhance this Socratic ideal of intellectual autonomy. Conversely, AI systems, with their ability to provide personalized learning paths and instant feedback, can potentially support student autonomy by allowing learners to progress at their own pace and explore topics of interest. However, there's a risk that over-reliance on AI recommendations could lead to a form of intellectual dependency, akin to the prisoners in Plato's allegory of the cave who mistake shadows for reality \cite{kraut2017plato}. The challenge, then, is to use AI as a tool for empowerment rather than as a crutch that limits independent thinking.

Aristotle's concept of \emph{eudaimonia}, or human flourishing, provides another valuable perspective on student autonomy in the age of AI. Aristotle believed that true happiness comes from realizing one's full potential through the exercise of virtues and the development of practical wisdom (phronesis) \cite{kristjansson2015aristotelian}. In an educational context, this translates to helping students develop not just knowledge and skills, but the ability to make sound judgments and take initiative in their learning. AI can support this goal by providing rich, interactive learning experiences and opportunities for self-assessment. For instance, AI-powered simulations can allow students to experiment with complex scenarios, developing their decision-making skills in a safe environment. However, it's crucial that these AI tools are designed to scaffold students towards greater autonomy, gradually reducing support as learners become more proficient, much like the process of habituation that Aristotle saw as key to developing virtues.

The ancient Greek concept of \emph{arete}, or excellence, is also relevant here. The Greeks believed that education should aim at developing excellence in all aspects of a person's life, not just in specific skills or knowledge areas \cite{jaeger1939ideals}. In the context of AI-enhanced education, this reminds us of the importance of fostering a holistic sense of initiative and autonomy that extends beyond the ability to navigate AI systems effectively. For example, while AI can efficiently deliver content and assess basic knowledge, it should also be leveraged to encourage students to set their own learning goals, design their own projects, and evaluate their own progress. This aligns with the Greek ideal of self-knowledge, epitomized by the Delphic maxim ``Know thyself,'' which was central to Socratic philosophy \cite{vlastos1991socrates}.

However, fostering student autonomy in an AI-enhanced environment is, again, not without challenges. One significant concern is the potential for AI systems to create a false sense of autonomy. If students rely too heavily on AI recommendations or automated learning paths, they may develop a form of learned helplessness, becoming less capable of independent decision-making and critical thinking. This echoes the above-mentioned discussion by Plato: just as Plato worried that people might mistake the ability to access written information for true wisdom, we must be cautious about conflating the ability to navigate AI systems with genuine intellectual autonomy.

Furthermore, the development of critical thinking skills becomes even more crucial in an AI-rich environment. Students need to be able to evaluate the information and recommendations provided by AI systems, understanding their limitations and potential biases. This critical stance towards AI aligns with the skeptical tradition in Greek philosophy, which emphasized the importance of questioning accepted beliefs and examining evidence \cite{guthrie1971sophists}.

In conclusion, while AI offers powerful tools for personalized learning and support, the ultimate goal of education – as envisioned by ancient Greek philosophers – remains the development of autonomous, critically thinking individuals capable of pursuing excellence and wisdom. By thoughtfully integrating AI in ways that support rather than supplant student initiative, and by explicitly teaching the skills needed for genuine autonomy in an AI-rich world, we can work towards an educational model that combines the best of ancient wisdom with modern technology.

\section{Discussion \& Conclusion}
As we look towards the future of Artificial Intelligence (AI) in education, we find ourselves at a crossroads reminiscent of the philosophical debates in ancient Greece. The potential for AI to revolutionize learning is immense, yet it brings with it profound ethical and practical challenges. By examining this future through the lens of ancient Greek philosophy, we can gain valuable insights into how to navigate this transformative period in education.

The promise of AI in education echoes Plato's vision of an ideal society in his Republic, where each individual receives an education tailored to their abilities and societal role \cite{kraut2017plato}. AI's capacity for personalization and adaptive learning could bring us closer to this ideal, offering each student a uniquely optimized educational experience. However, just as Plato's vision raised questions about equality and predetermination, we must critically examine the implications of AI-driven educational sorting and tracking.

Aristotle's concept of eudaimonia, or human flourishing, provides a valuable framework for considering the ultimate goals of AI in education \cite{kristjansson2015aristotelian}. From this perspective, the true measure of AI's success in education should not be merely efficiency or knowledge acquisition, but its ability to contribute to the holistic development of individuals. This aligns with the Greek ideal of paideia, which viewed education as a means of cultivating virtue and civic responsibility \cite{jaeger1939ideals}.

As we develop more sophisticated AI systems, we must be mindful of another Delphic maxim, ``Nothing in excess''. This principle of moderation, central to Greek philosophy, reminds us to strike a balance in our use of AI in education. While embracing the benefits of AI, we must be cautious not to diminish the crucial human elements of education – the Socratic dialogue, the mentorship, the collaborative exploration of ideas.

The potential for AI to enhance human capabilities in education resonates with the myth of Prometheus, who gave fire (technology) to humans. Just as this gift had transformative potential but also risks, AI in education offers great promise but requires wisdom in its application. We must strive to be, as Prometheus was, ``forethoughtful'' (which is the meaning of his name) in our approach to AI in education. However, the integration of AI in education also raises concerns that echo the skepticism of ancient Greek philosophers. The Sophists' emphasis on the potential for rhetoric to manipulate finds a modern parallel in concerns about AI's potential to influence or misinform. As Socrates critiqued the Sophists for prioritizing persuasion over truth \cite{guthrie1971sophists}, we must ensure that AI systems in education are designed to promote genuine understanding rather than mere performance on standardized metrics.

Moreover, as AI systems become more advanced, we face questions about the nature of knowledge and wisdom that would not be unfamiliar to ancient Greek thinkers. Plato's Theory of Forms distinguished between the physical world of appearances and the abstract world of ideas \cite{kraut2017plato}. In our context, we must consider: Can AI, which operates on data derived from the physical world, help students access deeper, abstract truths? Or is there a form of wisdom that remains uniquely human?

The Socratic method, with its emphasis on questioning and dialogue, remains a powerful model for the future of education, even as AI becomes more prevalent. AI systems should be designed not just to provide answers, but to stimulate curiosity and critical thinking in the Socratic tradition. The goal should be to create AI that acts as a partner in the educational process, enhancing rather than replacing the human-to-human dialogue that Socrates saw as central to learning.

Looking to the future, the development of AI in education must be guided by ethical principles. Here, we can draw inspiration from Aristotle's virtue ethics, which emphasized the importance of developing good character \cite{kristjansson2015aristotelian}. As we create and implement AI systems in education, we must consider not just their efficiency or effectiveness, but their impact on the moral and ethical development of students.

The Stoic philosophy, with its emphasis on focusing on what is within our control, offers a valuable perspective on how to approach the uncertainties of AI's future in education \cite{sellars2014stoicism}. While we cannot predict all the ways AI will transform education, we can control how we prepare for this future – by fostering adaptability, critical thinking, and ethical reasoning in both educators and students.

In conclusion, the future of AI in education holds great promise, but realizing this potential requires us to engage with fundamental questions about the nature of knowledge, learning, and human flourishing – questions that were central to ancient Greek philosophy. By maintaining a critical yet optimistic view, informed by the wisdom of classical thinkers, we can work towards an AI-enhanced educational future that amplifies human potential while preserving the essential humanistic values that have been at the heart of education for millennia.



\bibliographystyle{ACM-Reference-Format}
\bibliography{biblio}

\end{document}